\documentclass[prl,twocolumn,amsmath,amssymb]{revtex4-1}
\usepackage{bm}
\usepackage{graphicx}
\usepackage{color}
\usepackage{dcolumn,amsmath}
\usepackage[normalem]{ulem}

\begin{document}
\newcommand{\HH}{H$_2$}
\newcommand{\pHH}{\emph{para}-H$_2$}
\newcommand{\oHH}{\emph{ortho}-H$_2$}
\newcommand{\Eref}[1]{Eq.~(\ref{#1})}
\newcommand{\Fref}[1]{Fig.~\ref{#1}}
\newcommand{\etal}{\emph{et al.}}

\title{Unravelling the stereodynamics of cold HD-H$_2$ collisions}

\author{James F. E. Croft}
\author{Naduvalath Balakrishnan}
\affiliation{Department of Chemistry and Biochemistry, University of Nevada, Las Vegas, Nevada 89154, USA}
\author{Meng Huang}
\author{Hua Guo}
\affiliation{Department of Chemistry and Chemical Biology,
University of New Mexico,
Albuquerque, New Mexico 87131, USA}

\begin{abstract}
Measuring inelastic rates with partial wave resolution requires temperatures
close to a Kelvin or below, even for the lightest molecule.
In a recent experiment Perreault~\etal~\cite{perreault.mukherjee.ea:cold}
studied collisional relaxation of excited HD molecules in the $v=1, j=2$ state
by \emph{para}- and \oHH\ at a temperature of about 1~K,
extracting the angular distribution of scattered HD in the $v=1,j=0$ state.
By state-preparation of the HD molecules, control of the angular distribution
of scattered HD was demonstrated.
Here, we report a first-principles simulation of that experiment which enables
us to attribute the main features of the observed angular distribution to a
single $L=2$ partial-wave shape resonance.
Our results demonstrate important stereodynamical insights that can be
gained when numerically-exact quantum scattering calculations are combined
with experimental results in the few-partial-wave regime.
\end{abstract}
\maketitle

\section{Introduction}
The ultimate goal of chemistry is the complete quantum state control of both
reactants and products. Understanding the state-to-state stereodynamics of collision
processes is a perquisite for attaining such control
\cite{bernstein.herschbach.ea:dynamical,zare:laser,aldegunde.miranda.ea:how,perreault.mukherjee.ea:cold}.
Reducing the collision energy to a Kelvin or less simplifies collisional
processes by restricting the relevant number of partial waves.
Thanks to recent developments in molecule cooling and trapping
\cite{wynar.freeland.ea:molecules,regal.ticknor.ea:creation,
sawyer.lev.ea:magnetoelectrostatic,shuman.barry.ea:laser,hummon.yeo.ea:2d,
akerman.karpov.ea:trapping,anderegg.augenbraun.ea:radio,truppe.williams.ea:molecules}
and merged beams
\cite{henson.gersten.ea:observation,jankunas.bertsche.ea:dynamics,
klein.shagam.ea:directly,perreault.mukherjee.ea:supersonic}
it is now increasingly possible to study molecular systems in this
few-partial-wave regime
\cite{ospelkaus.ni.ea:quantum-state,knoop.ferlaino.ea:magnetically,
rui.yang.ea:controlled,perreault.mukherjee.ea:quantum,wolf.dei.ea:state-to-state,
poel.zieger.ea:cold,amarasinghe.suits:intrabeam}.

The stereodynamics of many inelastic and reactive molecular encounters is
strongly influenced by resonances, which occur via either tunneling through a
centrifugal barrier (shape resonance) or coupling to a bound state of a
closed channel (Fano-Feshbach resonance)
\cite{chandler:cold,klein.shagam.ea:directly,amarasinghe.suits:intrabeam,bergeat.onvlee.ea:quantum}.
Low-energy collisions of light molecules such as H$_2$ in the region of 1~K
occur in a ``Goldilocks zone" -- neither too hot nor too cold -- where
chemical processes are dominated by just a few partial waves.
However, experimental studies of molecular collisions and measurements of product
angular distributions in this regime have been a significant challenge,
in particular for neutral molecules such as H$_2$ and HD which are not
magnetically trappable and have zero or very small dipole moment (for HD).

In a landmark experiment, Perreault~\etal\ reported four-vector correlations
for collisions of excited HD molecules in the $v=1,j=2$ level with D$_2$ and
H$_2$ at a collision energy around 1~K
\cite{perreault.mukherjee.ea:quantum,perreault.mukherjee.ea:cold}.
In the experiment HD and H$_2$/D$_2$ are co-expanded in a single beam,
and the HD molecules are prepared in one of two specific well-defined states
using Stark-induced adiabatic Raman passage (SARP).
SARP combined with a co-expansion in a molecular beam therefore provides a
powerful tool for studying the stereodynamics of cold
collisions without having to explicitly remove their kinetic energy.

Here, we report a first-principles simulation of the experiment of
Perreault~\etal\ based on full-dimensional quantum scattering calculations.
In doing so we unravel the stereodynamics of the collision process
and attribute the observed experimental angular distribution to a $L=2$
shape resonance in the incoming channel.
We also explain the origin of the symmetric angular distribution observed in
the experiment.

\section{Methods}
Being the simplest neutral molecule-molecule system, H$_2$+H$_2$/HD collisions
are amenable to full-dimensional quantum scattering calculations
\cite{ying-lin.guo:full-dimensional,pogrebnya.clary:full-dimensional,
gatti.otto.ea:rotational,quemener.balakrishnan.ea:vibrational}
and high quality \emph{ab initio} potential energy surfaces
are available.
In this work we have used the full-dimensional \HH-\HH\ potential of Hinde~\cite{hinde:six-dimensional},
which has been used extensively in recent years to study
scattering of H$_2$ on H$_2$ and its isotopologs
\cite{santos.balakrishnan.ea:quantum,balakrishnan.quemener.ea:full-dimensional}.
Its features compare well with the other available potentials
for the H$_2$-H$_2$ system~\cite{boothroyd.dove.ea:accurate,patkowski.cencek.ea:potential}.
In particular, its accuracy is comparable to the four-dimensional potential of
Patkowski~\etal~\cite{patkowski.cencek.ea:potential} which is considered to
be the most accurate for the H$_2$-H$_2$ system (with an uncertainty
of about 0.15 K or about 0.3\% at the minimum of the potential well).

Scattering calculations for collisions of HD with \HH\ were performed
in full-dimensionality using a modified version of the TwoBC code~\cite{krems:twobc}.
The methodology is well established and outlined in detail
\cite{quemener.balakrishnan.ea:vibrational,quemener.balakrishnan:quantum,
santos.balakrishnan.ea:quantum},
and has been  applied to other similar systems
\cite{yang.zhang.ea:quantum,yang.zhang.ea:full-dimensional,
yang.wang.ea:full-dimensional,santos.balakrishnan.ea:vibration-vibration}.
Here we briefly review the methodology in order to define notation.
The scattering calculations are performed within the
time-independent close-coupling formalism yielding the usual asymptotic
$S$ matrix~\cite{arthurs.dalgarno:theory}.
For convenience, we label each asymptotic channel by the
combined molecular state (CMS) $\alpha=v_1 j_1 v_2 j_2$,
where $v$ and $j$ are vibrational and rotational quantum numbers respectively
and the subscript 1 refers to HD and 2 to \HH.
The integral cross section for state-to-state rovibrationally inelastic
scattering is given by,
\begin{eqnarray}
\sigma_{\alpha \to \alpha'} = &&\frac{\pi}{(2j_1+1)(2j_2+1)k_\alpha^2} \\
&\times& \sum_{J,j_{12},j'_{12},l,l'}(2J+1) |T^{J}_{\alpha lj_{12},\alpha'l'j'_{12}}|^{2}. \nonumber
\end{eqnarray}
where $k^2=2 \mu E/\hbar^2$, $T^J = 1-S^J$, $L$ is the orbital angular momentum,
$J$ the total angular momentum ({\bf J} = {\bf L} + {\bf j$_{12}$}),
and {\bf j$_{12}$} = {\bf j}$_1$ + {\bf j}$_2$.
To compute the differential cross sections relevant to this work
we also need the scattering amplitude, which has previously been given by Schaefer
\etal\ ~\cite{schaefer.meyer:theoretical} in the helicity representation,
\begin{eqnarray}
q = \frac{1}{2k_\alpha}\sum_{J}&(2J+1)&\sum_{j_{12},j'_{12},l,l'}i^{l-l'+1}T^J_{\alpha lj_{12},\alpha'l'j'_{12}}d^J_{m_{12},m'_{12}}(\theta) \\
&\times& \big\langle j'_{12}m'_{12}J-m'_{12}|l'0\big\rangle \big\langle j_{12}m_{12}J-m_{12}|l0\big\rangle \nonumber\\
&\times& \big\langle j'_1m'_1j'_2m'_2|j'_{12}m'_{12}\big\rangle \big\langle j_1m_1j_2m_2|j_{12}m_{12}\big\rangle \nonumber
\end{eqnarray}
where $d^J_{m_{12},m'_{12}}(\theta)$ is Wigner's small $d$ rotation matrix. 
The rovibrational state-to-state differential cross section is then given by
\begin{eqnarray}
\label{eqn:dcs}
\frac{d\sigma_{\alpha \to \alpha'}}{d\Omega} &&= \frac{1}{(2j_1+1)(2j_2+1)} \\
&\times&\sum_{m_1,m_2,m_{12},m'_1,m'_2,m'_{12}} |q_{\alpha,m_1,m_2,m_{12} \to \alpha',m'_1,m'_2,m'_{12}}|^{2}. \nonumber
\end{eqnarray}

\section{Results}
\begin{figure}[tb]
\centering
\includegraphics[width=0.97\columnwidth]{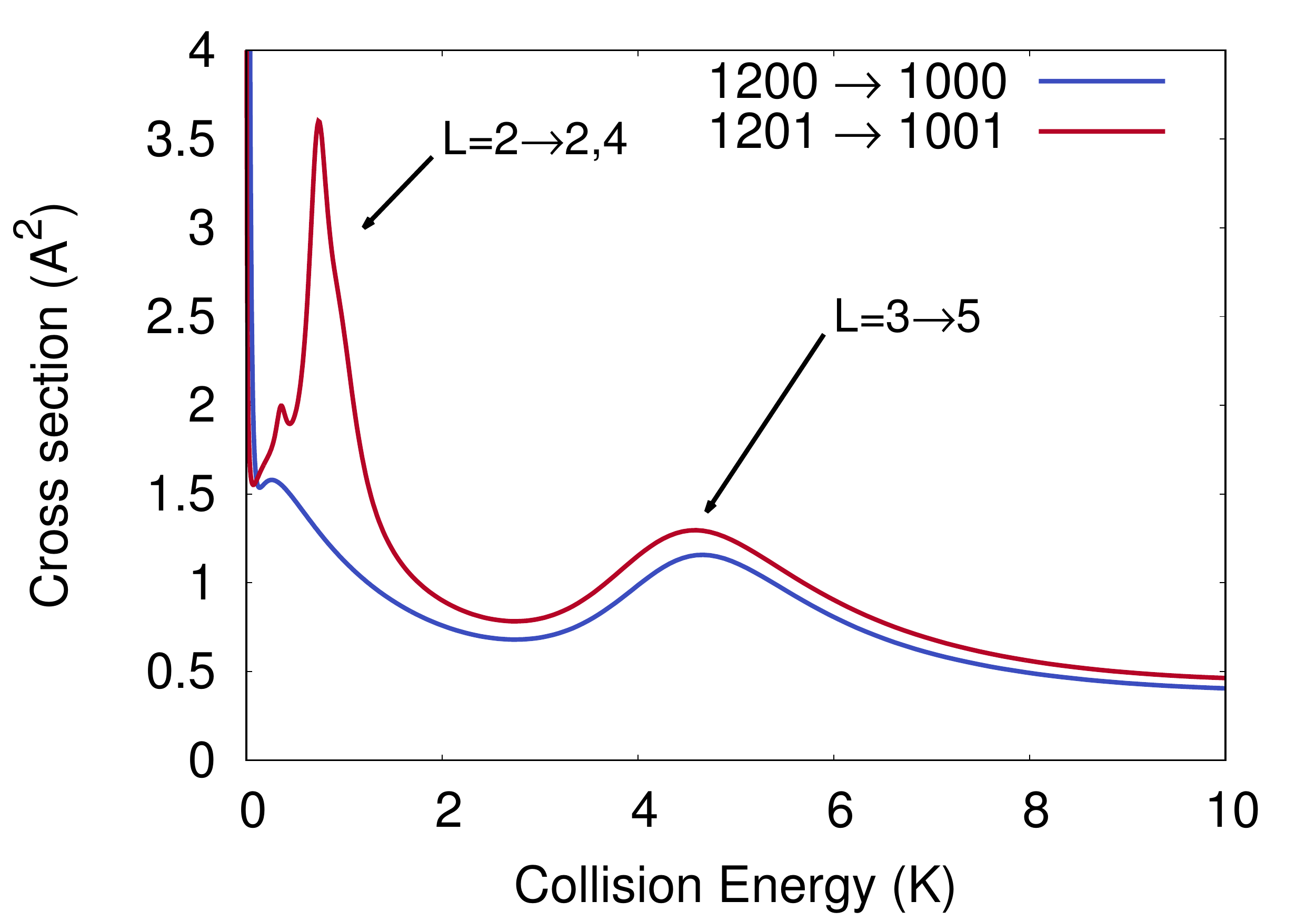}
\caption{Integral state-to-state cross sections for HD($v=1,j=2$) $\to$
HD($v=1,j=0$) in collisions with \HH($j=0,1$)}
\label{fig:ics}
\end{figure}

In the recent work of Perreault \etal\ collisions of HD($v=1, j=2$) with
H$_2$($v=0, j=0,1$) were studied in the 0-10~K regime and the angular distribution
of HD($v=1, j=0$) measured~\cite{perreault.mukherjee.ea:cold}.
Figure~\ref{fig:ics} shows the corresponding theoretical integral cross section for,
$\alpha=1200 \to 1000$ and $\alpha=1201 \to 1001$. 
It is clearly seen that there are shape resonances for collisions with
both \oHH\ and \pHH, in the vicinity of 1~K, with the dominant feature
being a $L=2$ shape resonance with \oHH\ at around 1~K.

\begin{figure}[tb]
\centering
\includegraphics[width=0.97\columnwidth]{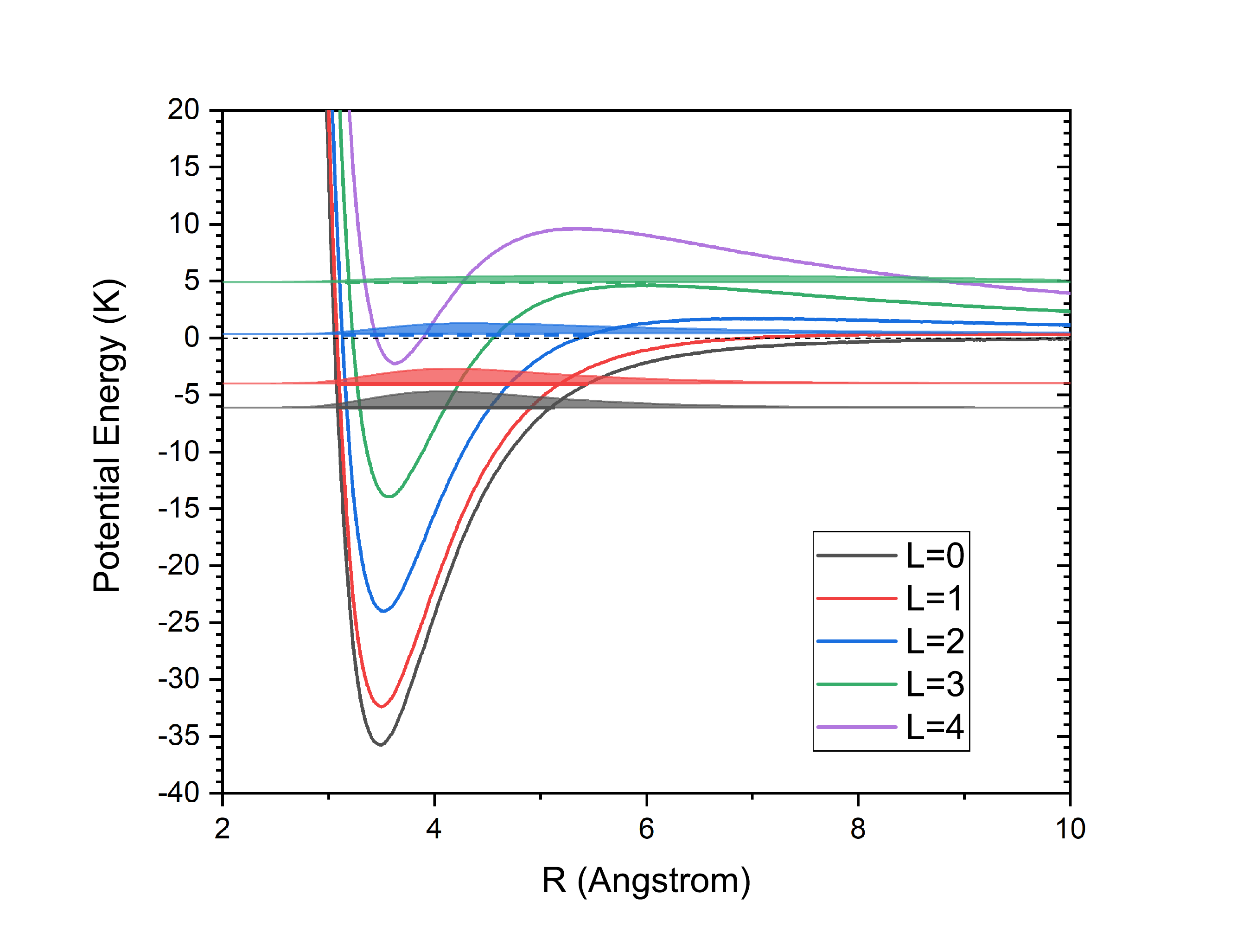}
\caption{One-dimensional adiabatic potentials and wavefunctions of the HD-H$_2$
system as a function of $R$.}
\label{fig:eff}
\end{figure}

In order to gain insight into the nature of the resonances seen in~\Fref{fig:ics}
we analyzed the effective potentials corresponding to different incoming partials waves $L$,
\begin{eqnarray}
V^J(R)&=&\epsilon_{v_1j_1v_2j_2} \\  \nonumber
&+&U^{J}_{v_1j_1v_2j_2 L j_{12}, v'_1 j' _1 v'_2 j'_2 L' j'_{12}}(R) + \frac{L(L+1)\hbar^2}{2\mu R^2}.
\end{eqnarray}
The first term is the energy of the CMS obtained by adding the asymptotic
rovibrational energies of HD and \HH .
The second term is the diabatic potential energy coupling matrix and the
third term is the centrifugal potential for the orbital angular momentum $L$.
At large intermolecular separations, the energies of the different channels
that correspond to the same CMS converge to its asymptotic value.
The effective potential matrix is diagonalized at each value of $R$ and the
eigenvalues as a function of $R$ correspond to a series of adiabatic potentials.
Bound or quasibound states of these one-dimensional potentials
correspond to HD-H$_2$ complexes, and the decay of the quasibound states leads
to the resonances seen in~\Fref{fig:ics}.
Figure~\ref{fig:eff} shows the potentials for $L =0,1,2,3,4$ for the asymptotic
state 1201 along with the corresponding one-dimensional wavefunctions -- shown
at the bound or quasibound energies.
It is the quasibound states at $\approx 1$~K and $\approx 5$~K in the $L=2$ and 3
channels respectively which lead to the shape resonances seen in~\Fref{fig:ics}.
The corresponding outgoing dominant partial waves are $L'=2$ and 4 for $L=2$
and $L'=5$ for $L=3$ as shown in~\Fref{fig:ics}.

The experimental setup is described in detail in a series of papers by
Perreault \etal~\cite{perreault.mukherjee.ea:quantum,perreault.mukherjee.ea:supersonic,perreault.mukherjee.ea:cold}.
Here we only outline the details necessary for making a comparison with
our theory results.
In the experiment HD and \HH\ are co-expanded in a single beam.
The HD molecule is prepared in one of two specific states using
the SARP technique.
H-SARP prepares the HD($v_1=1, j_1=2$) in a state $|j_1=2,m_1=0\rangle$,
where $m_1$ refers to the angular-momentum component along the relative
velocity axis, in which case the HD bond is aligned parallel to the relative velocity.
V-SARP, prepares the HD($v_1=1, j_1=2$) in a state
\begin{eqnarray}
\sqrt{\frac{3}{8}}|j_1=2,m_1=-2\rangle&-&\frac{1}{2}|j_1=2,m_1=0\rangle \\
&+&\sqrt{\frac{3}{8}}|j_1=2,m_1=2\rangle, \nonumber
\end{eqnarray}
in which case the HD bond is aligned perpendicular to the relative velocity.
The H and V in H-SARP and V-SARP refer to the horizontal and vertical orientations
of the SARP laser relative to the beam velocity.
The \HH\ on the other hand is not state prepared and the ratio of \pHH\ to
\oHH\ in the beam is taken to be 1 to 3.
The experiment then measures the rate of HD($v_1=1, j_1=0$) scattered into a solid angle
$\Omega$ relative to the beam velocity.

\begin{figure}[t]
\centering
\includegraphics[width=0.97\columnwidth]{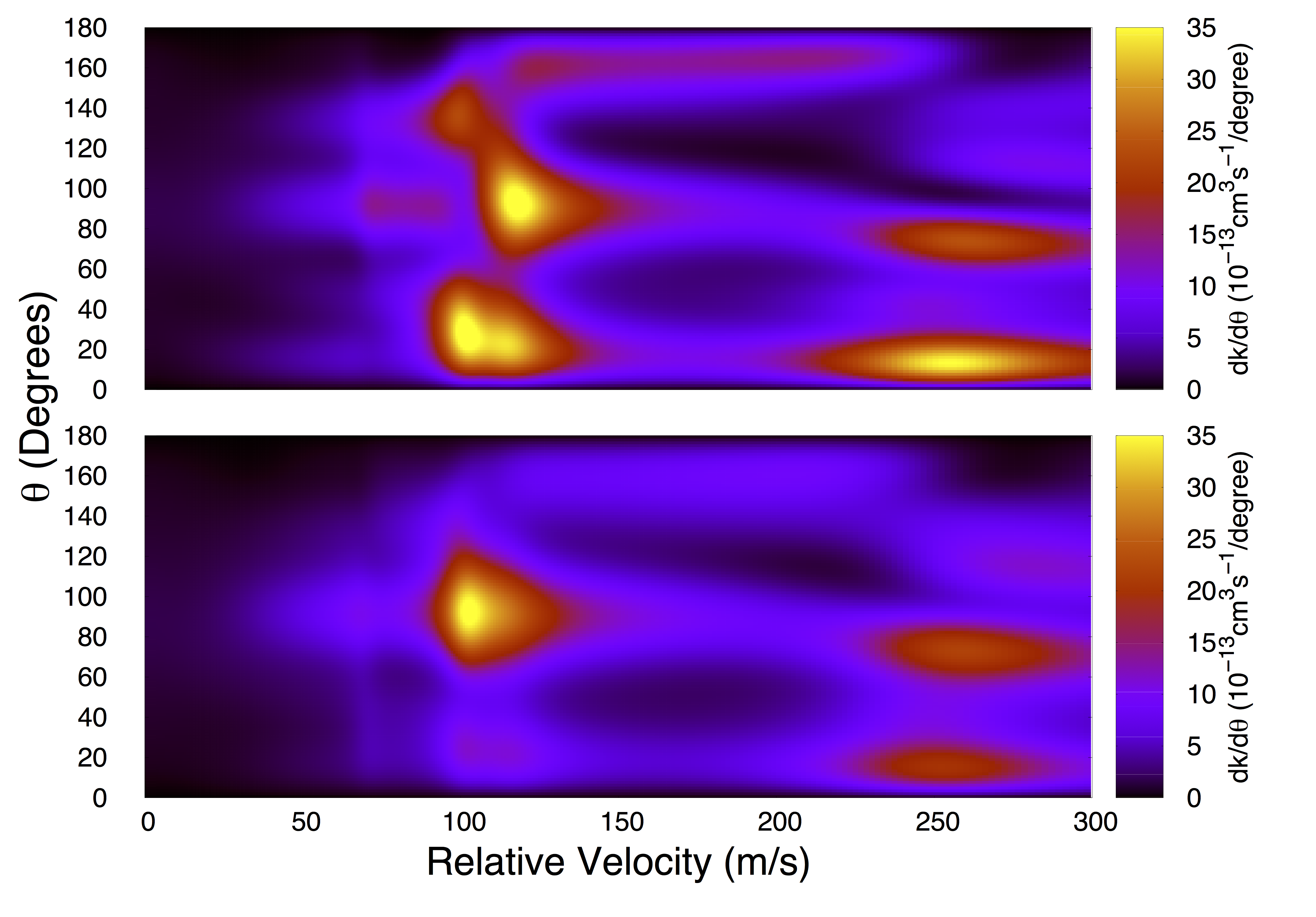}
\caption{The differential state-to-state rate for the transition
HD($v=1,j=2$) $\to$ HD($v=1,j=0$) in collisions with \oHH\
where the HD was prepared with H-SARP (upper panel)  and V-SARP (lower panel).}
\label{fig:kcs}
\end{figure}

In order to compare with the experimental result we need to account for these
experimental particulars.
When molecules are prepared using H-SARP or V-SARP~\Eref{eqn:dcs} for the
differential cross-section has to be modified to account for the interference
between the different $m$'s in the initial state preparation.
For H-SARP it becomes
\begin{eqnarray}
\frac{d\sigma^H_{\alpha \to \alpha'}}{d\omega} &=& \frac{1}{(2j_2+1)} \\
&&\sum_{m_{12},m'_{12},m_1',m_2,m'_2} |q_{\alpha,m_1=0,m_2,m_{12} \to \alpha',m'_1,m'_2,m'_{12}}|^{2}, \nonumber
\label{eqn:dcs_hsarp}
\end{eqnarray}
while for V-SARP it becomes
\begin{eqnarray}
\frac{d\sigma^V_{\alpha \to \alpha'}}{d\omega} = \frac{1}{(2j_2+1)} \\
\sum_{m_{12},m'_{12},m_1',m_2,m'_2} &|&\sqrt{\frac{3}{8}}q_{\alpha,m_1=-2,m_2,m_{12} \to \alpha',m'_1,m'_2,m'_{12}} \nonumber\\
&-&\frac{1}{2}q_{\alpha,m_1=0,m_2,m_{12} \to \alpha',m'_1,m'_2,m'_{12}} \nonumber \\
&+&\sqrt{\frac{3}{8}}q_{\alpha,m_1=+2,m_2,m_{12} \to \alpha',m'_1,m'_2,m'_{12}}|^{2}. \nonumber
\label{eqn:dcs_vsarp}
\end{eqnarray}
As seen in~\Fref{fig:ics} the dominant feature seen in the experiment is
expected to be an $L=2$ shape resonance from collisions with \oHH,
especially when the relative population of \oHH\ and \pHH\ in the beam
is taken into account.
Figure~\ref{fig:kcs} shows the differential rate (defined below) as a function of
the relative velocity for the  state-to-state transition, HD($v=1,j=2$) $\to$
HD($v=1,j=0$) in collisions with \oHH\ for H-SARP and V-SARP\@.
The $L=2$ shape resonance seen in~\Fref{fig:ics} is clearly visible
at around 100~ms$^{-1}$ ($\approx$1~K).
The initial alignment of the HD with respect to the beam velocity clearly makes
a significant difference in the angular distribution.
For V-SARP, where the HD bond axis is aligned perpendicular to the beam axis,
the dominant scattering is at around 90 degrees whereas for H-SARP,
where the HD bond axis is aligned parallel to the beam axis,
there is also significant forward scattering at around 20 degrees.
The equivalent figures for collisions with \pHH\ are given in the
supplemental materials.

In order to make an explicit comparison with the experimental angular distribution,
we also have to average over both the relative velocity distribution and the
relative populations of \oHH\ and \pHH.
The experimental velocity distributions for HD and \HH\ are given by the Gaussian
distributions $P(v_{\mathrm{HD}}) \sim \mathcal N(\mu_{\mathrm{HD}}=2814, \sigma_{\mathrm{HD}}^2=71^2/2)$ and
$P(v_{\mathrm{H}_2}) \sim \mathcal N(\mu_{\mathrm{H}_2}=2740, \sigma_{\mathrm{H}_2}^2=105^2/2)$,
where $v$, $\mu$, and $\sigma$ are in units of ms$^{-1}$~\cite{perreault.mukherjee.ea:supersonic}.
With the relative velocity defined as $v_{\mathrm{rel}} = v_{\mathrm{HD}} - v_{\mathrm{H}_2}$ the
relative velocity distribution is then given by convolving the two distributions
yielding
$P(v_{\mathrm{rel}}) \sim \mathcal N(\mu_{\mathrm{rel}} = \mu_{\mathrm{HD}}-\mu_{\mathrm{H}_2}, \sigma_{\mathrm{rel}}^2=\sigma_{\mathrm{HD}}^2+\sigma_{\mathrm{H}_2}^2$).
In the experiment the scattering angle $\theta_{\mathrm{exp}}$ is defined relative to the
beam velocity, therefore for positive relative velocities (HD catching up with \HH)
$\theta_{\mathrm{exp}} = \theta$ whereas for negative relative velocities
(HD being caught up by \HH) $\theta_{\mathrm{exp}} = \pi - \theta$.
The velocity averaged differential rate, for \emph{ortho}- or \pHH,
is therefore given by
\begin{eqnarray}
\frac{dk(\theta_{\mathrm{exp}})}{d\theta_{\mathrm{exp}}} &=& \int_{-\infty}^{0} |v_{\mathrm{rel}}|\frac{d\sigma(\pi-\theta)}{d\theta}P(v_{\mathrm{rel}}) dv_{\mathrm{rel}} \nonumber \\
 &+& \int_{0}^{\infty} |v_{\mathrm{rel}}|\frac{d\sigma(\theta) }{d\theta}P(v_{\mathrm{rel}}) dv_{\mathrm{rel}},
\label{eqn:rel_vel}
\end{eqnarray}
by weighting them with the experimental population of \emph{para}- and \oHH\
(25\% and 75\% respectively) a direct comparison can be made with experiment.

\begin{figure}[t]
\centering
\includegraphics[width=0.97\columnwidth]{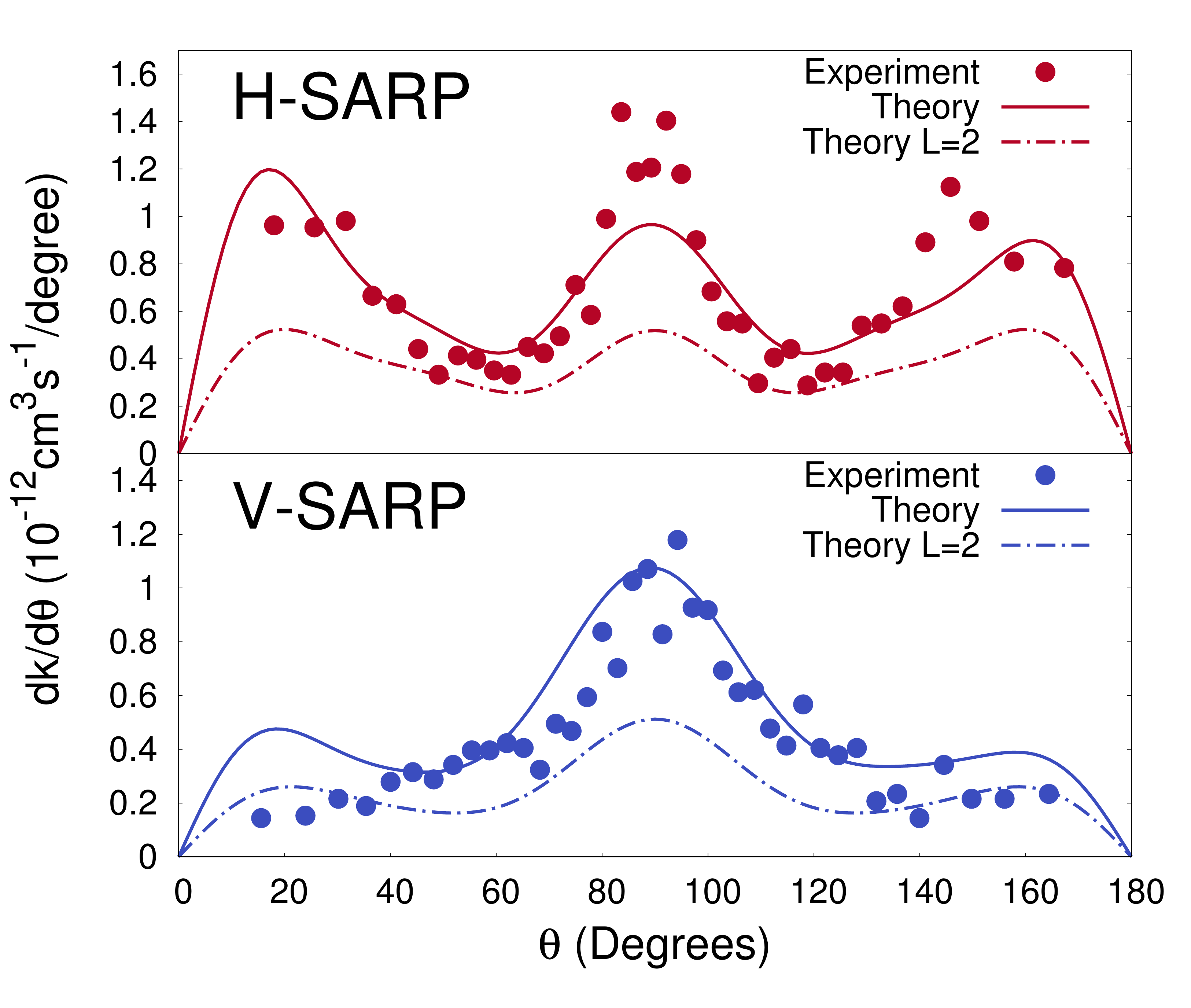}
\caption{The velocity averaged differential state-to-state rate for
HD($v=1,j=2$) $\to$ HD($v=1,j=0$) in collisions with \pHH\ and \oHH\ for
HD prepared using H-SARP and V-SARP\@.
The solid dots are the corresponding experiment results of Perreault
\etal~\cite{perreault.mukherjee.ea:cold}.}
\label{fig:exp}
\end{figure}

Figure~\ref{fig:exp} compares our theory results with the experimental
data presented in Perreault \etal~\cite{perreault.mukherjee.ea:cold}.
The experimental results for both H-SARP and V-SARP have been scaled by the same
factor (0.009).
It is seen we find excellent agreement with the experimental results
capturing the main features, as well as getting the relative magnitude
of H-SARP and V-SARP correct.
We note that this means we also get agreement with the higher integral
rate reported for H-SARP compared to V-SARP.
Comparing~\Fref{fig:exp} with~\Fref{fig:kcs} we are able to attribute the
observed features to a specific resonance.
This is especially clear in the case of V-SARP where the strong central feature
is clearly due to the $L=2$ shape resonance found at 100~ms$^{-1}$.
The $L=2$ contribution for collisions with \oHH\ is explicitly shown in~\Fref{fig:kcs}
as dashed lines, which can be seen to make up over half of the observed rate as
well as giving the overall form to the angular distribution.
In the case of H-SARP however there is a backwards scattering feature (at around 160~degrees)
seen in the experiment which is not present in the theoretical result. 
This apparent backwards scattering is in fact an artifact of the velocity
averaging of~\Eref{eqn:rel_vel} and is actually forward scattering
of HD from collisions with negative relative velocities.
More generally the approximate symmetry of the measured angular distribution seen here
is a direct consequence of the approximate symmetry of the relative velocity
distribution of this kind of experimental setup, which leads to nearly equal
contributions from positive and negative relative velocities in~\Eref{eqn:rel_vel}.
The separate contributions to the angular distribution from positive and negative
velocities are given in the supplemental materials.
We are therefore able to unambiguously attribute the observed feature to an
$L=2$ shape resonance for collisions of HD($v=1, j=2$) with H$_2$($j=1$).
We note that there is also a large $L=2$ shape resonance
for collisions of HD($v=0, j=2$) with H$_2$($j=0$) between 0.1 and 1~K
which disappears for HD($v=1$).
If this resonance is also present for HD($v=1$), say if the potential well were
actually slightly deeper, it would not change this conclusion as it would only affect
the overall magnitude of the cross section but not its form (we have checked this
explicitly by computing the HD($v=0, j=2$) $\to$ HD($v=0, j=0$) cross sections).

\section{Conclusions}
We have performed numerically-exact quantum scattering calculations for low
energy collisions of quantum-state prepared HD with \HH{},
finding excellent agreement with experiment for the angular distribution
of scattered HD.
Our computations provide a complete numerical simulation of the experiment with
full quantum-state resolution, including, orientation of the HD molecule
relative to the molecular beam axis.
We were able to unravel the stereodynamics of the collision process and
attribute the observed angular distribution to a single $L=2$ shape resonance
in the incoming channel.
This demonstrates the enormous potential of low energy beam experiments
for studying inelastic collision processes at the single partial wave level,
and the unique insights that can be gained in the collision dynamics when
combined with numerically-exact scattering calculations.
The excellent agreement between theory and experiment for this benchmark
system also provides an independent confirmation of the accuracy of the
H$_2$-H$_2$ interaction potential for collisional studies near 1~K,
a regime also of significant interest in astrophysics.

\section*{Acknowledgments}
We acknowledge support from the  US Army Research Office,
MURI grant No.~W911NF-12-1-0476 (N.B.),
the US National Science Foundation, grant No.~PHY-1505557 (N.B.),
and Department of Energy, grant number DE-SC0015997 (H. G.).
We thank Dick Zare, Nandini Mukherjee, and William Perreault for many
stimulating discussions and for sharing their experimental data.
\bibliography{../../all}

\end{document}